\documentclass[12pt]{article}
\usepackage[dvips]{graphicx}
\begin{document}
\parindent 2em

\begin{center}
\vspace{12mm}
{\LARGE Large-N theory of
strongly commensurate dirty bosons: absence of transition in two dimensions}
\vspace{15mm}

Matthew J. Case and Igor F. Herbut\\

{\it Department of Physics, Simon Fraser University\\
Burnaby, British Columbia, Canada\\V5A 1S6\\}

\end{center}
\vspace{10mm}

\begin{abstract}
The spherical limit of strongly commensurate dirty bosons is studied
perturbatively at weak disorder and numerically at strong disorder in
two dimensions (2D). We argue that disorder is not perfectly screened by
interactions, and consequently that
the ground state in the effective Anderson
localisation problem always remains localised. As a result there is 
only a gapped Mott insulator phase in the theory.
Comparisons with other studies and the parallel with disordered fermions
in 2D are discussed. We conjecture that while for the physical cases $N=2$
(XY) and $N=1$ (Ising) the theory should have the ordered phase, it may 
not for $N=3$ (Heisenberg). 
\end{abstract}

\vspace{5mm}
 
\noindent

\section{Introduction}

The problem of interacting bosons in a random potential is a
paradigmatic case of an interacting disordered system, and as such
has attracted much attention throughout the years \cite{hertz},
\cite{fisher}. Although in one version or another it has been used to
describe numerous physical situations \cite{herbut}, it has proven 
very difficult for a theoretical analysis, since it inextricably combines
the effects of interactions and Anderson localisation. Just like its
fermionic cousin the metal--insulator transition \cite{belitz}, the
problem of dirty bosons seems to lack a simple analytic mean-field
theory around which to begin a systematic study. Most of the information
on the dirty boson quantum phase transitions derive therefore from
numerical studies \cite{wallin}, and more recently from an
expansion around the lower critical dimension \cite{herbut1}.

In this paper we will be concerned with a limited class
of the dirty boson models at a commensurate filling, and study the
limit where the number of bosonic species $N$ is large \cite{tu}. As is well known
in this limit the mean-field theory, or the saddle-point approximation,
becomes the exact solution. The quantum mechanical action at $T=0$ that
defines our problem is 
\begin{equation}
\label{S}
S\left[\Psi\right] = \displaystyle\int {\rm d}^D\vec{x}{\rm d}\tau
\left\{
(\partial_\tau\Psi(\vec{x}, \tau))^2 + (\nabla\Psi(\vec{x}, \tau))^2 +
\left(V(\vec{x}) - \mu\right)\Psi^2(\vec{x}, \tau) + \frac{\lambda}{N}\Psi^4(\vec{x},
\tau)
\right\},
\end{equation}
where $\Psi(\vec{x}, \tau)$ is a real $N$-component bosonic field, 
$\Psi^2 = \displaystyle\sum_{\alpha = 1}^{N}\Psi_\alpha^2$, and $V(\vec{x})$
is a random (in space) external potential. For simplicity,
it will be assumed that $V(\vec{x})$ is uncorrelated,  so that
$\langle V(x) V(y)\rangle = W \delta(x-y)$. We will mostly be interested
in two dimensions $(D=2)$, but will leave a general $D$
in the action to comment later on results in other dimensions.
Note that disorder is assumed to be a random function only
of spatial coordinates, while it is completely correlated in (i.e.
independent of) the imaginary time. This is what makes it much stronger
than in the corresponding problem in classical mechanics. The theory (1)
for $N=2$ describes the superfluid order parameter in the Bose-Hubbard model,
at a density of bosons commensurate with the lattice \cite{fisher},
\cite{remark}, also known in literature as the random-rod problem
\cite{dorogovtsev}. For $N=3$ the theory may be used to describe disordered
quantum rotors, i.e. the magnetic quantum phase transitions in the
Heisenberg universality class in the presence of quenched randomness
\cite{sachdev}. When $N=1$ the theory describes a random system with the
Ising symmetry.  In general the action (1) provides a minimal description
of the quantum disordered interacting system, and for $N=\infty$ has been
studied by renormalisation group methods in the past \cite{kim,
hastings} with conflicting results. The purpose of this
paper is to shed some light on the physics implicit in this model, and,
in particular, to argue that the model allows no superfluid
phase in $D=2$.

To see what is involved in solving the problem in the (spherical) limit
$N=\infty$, perform the standard 
Hubbard--Stratonovich transformation on the quartic term and integrate out
all but one of the bosonic fields.  This leaves one with the
transformed action:
\begin{eqnarray}
\label{Seff}
S_{{\rm eff}}\left[\chi, \psi\right] & = & \displaystyle\int 
{\rm d}^D\vec{x}{\rm d}\tau\Bigg\{-\frac{N}{4\lambda}\chi^2(\vec{x}, \tau) +
(\partial_\tau\Psi_1(\vec{x}, \tau))^2 \\ & & + (\nabla\Psi_1(\vec{x}, \tau))^2
+ \left(V(\vec{x})
 + \chi(\vec{x},\tau) -
\mu\right)\Psi_1^2(\vec{x}, \tau)\Bigg\} \nonumber \\
& & + \frac{1}{2} (N-1)\ln\det\left\{-\partial_\tau^2 - \nabla^2 +
V(\vec{x}) + \chi(\vec{x},\tau) - \mu\right\}, \nonumber
\end{eqnarray}
which is just the original problem rewritten exactly. Assuming that the 
Hubbard--Stratonovich field at the saddle-point is independent of
imaginary time,
$\chi(\vec{x}, \tau) = \chi(\vec{x})$, and that $\Psi_1 (\vec{x}, \tau) =
N^{\frac{1}{2}}c\phi_0(\vec{x})$, the saddle-point equations become
\begin{equation}
\label{SP1}
\chi(\vec{x}) = \lambda\left<\vec{x},
\tau\left|\displaystyle\frac{1}{-\partial_\tau^2 - \nabla^2 + V(\vec{x})
+ \chi(\vec{x}) - \mu}\right|\vec{x}, \tau\right> + c^2\phi_0^2(\vec{x}),
\end{equation}
\begin{equation}
\label{SP2}
\varepsilon_0c = 0,
\end{equation}
where $\phi_\alpha(\vec{x})$ are the random eigenstates, and
$\varepsilon_\alpha$ the random eigenvalues of the susceptibility matrix
\begin{equation}
\label{SP3}
\left(-\nabla^2 + V(\vec{x}) + \chi(\vec{x}) - \mu\right)\phi_\alpha(\vec{x}) =
\varepsilon_\alpha\phi_\alpha(\vec{x}),
\end{equation}
with $\varepsilon_0$ being the lowest eigenvalue.
The Eqs. (\ref{SP1})-(\ref{SP3}) are completely standard, and the only
novelty compared to the case without disorder \cite{zinn} is the
random spectrum instead of the usual plane waves.  In principle,
one may expect to have two phases: 
$\varepsilon_0 \neq 0$ and $c = 0$, corresponding to the gapped
Mott insulator (MI), or
$\varepsilon_0 = 0$ and $c \neq 0$ which would represent a superfluid (SF).
The gapless insulating Bose-glass (BG) between the MI and the SF \cite{fisher}
should in general be absent, as we argue below.

With $V(\vec{x})=0$, the solution $\chi(\vec{x})=\chi_0$ is uniform, and the
model leads to the well-known large-N critical behavior in $D+1$ dimensions
\cite{zinn}. The correlation length exponent, for example, in the
pure case is $\nu=1/(D-1)$, and in $D=2$ the Harris criterion
\cite{harris} (that says that disorder is irrelevant if $\nu D >2$)
implies that disorder is precisely {\it marginal}. When $V(\vec{x})\neq 0$,
in the MI phase the saddle-point Eq. (\ref{SP1}), after integration over the
frequency, can be written in the basis $\{ \phi_\alpha \}$ as:
\begin{equation}
\label{SP1'}
\chi(\vec{x}) = \lambda\sum_{\alpha} \frac{\left|\phi_\alpha(\vec{x})\right|^2}
{\sqrt{\varepsilon_\alpha}}. 
\end{equation}
The functions $\phi_\alpha$ are the eigenstates of the {\it screened}, but
nevertheless random, potential $V(\vec{x}) + \chi(\vec{x})$, and would therefore
naively all be expected to be localised in $D=2$ \cite{abrahams}. 
In particular, for the localised ground state the first term in the sum
in Eq. (\ref{SP1'}) becomes large as $\varepsilon_0 \rightarrow 0$ precisely
in the region of localisation, which by self-consistency
implies that $\chi(\vec{x})$ is also large there. That, on the other hand,
then implies $\varepsilon_0$ is large, and not small as assumed, and one runs into
a contradiction. Evidently, for the spectrum to extend all the way
to zero the discrete sum in the last equation must be approximable
by an integral, so that the infrared singularity becomes integrable.
For this to occur the weight of each of the terms
corresponding to the low-energy states in Eq. (\ref{SP1'}) must vanish in the
thermodynamic limit as the inverse of the system size, which is tantamount to
delocalisation of the low-energy eigenstates. Put differently, 
the collapse of the gap must be accompanied by the
simultaneous delocalisation of the ground state, so that the
gapless phase is necessarily a SF. There can be no intermediate
localised BG in the model at $N=\infty$.

With this picture in mind the appearance of the superfluid phase in the
large-N model in $D=2$ appears rather counterintuitive: although
screening introduces correlations into the effective random potential, the
states should nevertheless always remain localised. In the rest 
of the paper we first show that although to the lowest order screening
does reduce the random potential, it does not make it completely smooth
and consequently the MI gap can not close. This conclusion is further
corroborated
by the numerical solution of the self-consistent equations on a lattice
and absence of the finite-size scaling of the gap and the ground state
participation ratio. In the closing section we compare our result
with other studies and speculate on the implications for physical
cases $N=1,2,3$. 

\section{Weak-disorder expansion}

\begin{figure}[h]
\begin{center}
\includegraphics{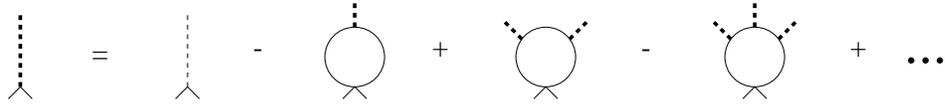}
\end{center}
\caption{Diagrammatic representation of Eq. (\ref{WDE}).  The heavy dashed line
represents the self-consistently screened random potential, while the thin dashed line is
the bare random potential.}
\label{Vdiag}
\end{figure}

For a given random configuration the self-consistent equations can not be
solved analytically, and one has to resort to numerical computations.
For weak disorder, however, we can expand the
matrix element in (\ref{SP1}) in powers of the {\it screened} potential.
To that end write $\chi(\vec{x}) = \chi_0 + \chi_1(\vec{x})$, where 
$\int\chi_1(\vec{x}){\rm d}^D\vec{x} = 0$. The uniform part $\chi_0$ is just the
renormalisation of the chemical potential, while 
$\widetilde{V}(\vec{x}) \equiv V(\vec{x}) +
\chi_1(\vec{x})$ is the screened potential, which should vanish with
vanishing randomness. Expanding the right hand side of (\ref{SP1}) in the MI phase in
$\widetilde{V}(\vec{x})$ and taking the Fourier transform, we get (for $q\neq 0$)
\begin{eqnarray}
\label{WDE}
\widetilde{V}(\vec{q}) & = & V(\vec{q}) -\lambda\Pi(\vec{q})\widetilde{V}(\vec{q})
+ \lambda\displaystyle\int{\rm d}\vec{k}
I_1(\vec{k}, \vec{q})\widetilde{V}(\vec{k})\widetilde{V}(\vec{q}-\vec{k}) \\
& & - \lambda\displaystyle\int{\rm d}\vec{k}{\rm d}\vec{l} I_2(\vec{k}, \vec{l},
\vec{q})\widetilde{V}(\vec{k})\widetilde{V}(\vec{l})\widetilde{V}(\vec{q}-\vec{k}-\vec{l}) +
{\cal O}(\widetilde{V}^4), \nonumber
\end{eqnarray}
where
\begin{equation}
\label{Pi}
\Pi(\vec{q}) \equiv \displaystyle\int{\rm d}\vec{p}{\rm d}\omega G_0(\omega,
\vec{p})G_0(\omega, \vec{p}+\vec{q}),
\end{equation}
is the standard polarisation bubble, and 
\begin{equation}
\label{I1}
I_1(\vec{k}, \vec{q}) \equiv \displaystyle\int{\rm d}\vec{p}{\rm d}\omega G_0(\omega,
\vec{p})G_0(\omega,
\vec{p}+\vec{k})G_0(\omega, \vec{p}+\vec{q}),
\end{equation}
and
\begin{equation}
\label{I2}
I_2(\vec{k}, \vec{l}, \vec{q}) \equiv \displaystyle\int{\rm d}\vec{p}{\rm d}\omega
G_0(\omega, \vec{p})G_0(\omega,
\vec{p}+\vec{k})G_0(\omega, \vec{p}+\vec{k}+\vec{l})G_0(\omega, \vec{p}+\vec{q}).
\end{equation}
The propagator for the clean case is given by $G_0^{-1}(\omega, \vec{p}) = \omega^2 + p^2
+ \Omega^2$, where $\Omega^2 \equiv \chi_0 - \mu > 0$ and is the MI gap.
Eq. (\ref{WDE}) can be represented diagrammatically as in Fig. \ref{Vdiag}.

We next
introduce the two point correlator
$\widetilde{W}(\vec{q})\delta(\vec{r}) =
\left<\widetilde{V}(\vec{q})\widetilde{V}(-\vec{q}+\vec{r})\right>$,
where $\left<\cdots\right>$ represents disorder averaging, as a measure of the
screened disorder. From Eq. (\ref{WDE}) it follows that
\begin{eqnarray}
\label{WDEW}
\widetilde{W}(\vec{q})\left\{1 + \lambda\Pi(\vec{q})\right\}^2 & = & W(\vec{q}) +
2\lambda\displaystyle\int{\rm
d}\vec{k}I_1(\vec{k}, \vec{q})\left<V(-\vec{q})\widetilde{V}(\vec{k})\widetilde{V}
(\vec{q}-\vec{k})\right> \\
& & -2\lambda\displaystyle\int{\rm d}\vec{k}{\rm d}\vec{l}I_2(\vec{k}, \vec{l},
\vec{q})\left<V(-\vec{q})\widetilde{V}(\vec{k})\widetilde{V}
(\vec{l})\widetilde{V}(\vec{q}-\vec{k}-\vec{l})\right> \nonumber \\
& & +\lambda^2\displaystyle\int{\rm d}\vec{k}{\rm d}\vec{k}'I_1(\vec{k},
\vec{q})I_1(\vec{k}', -\vec{q})\left<\widetilde{V}(\vec{k})\widetilde{V}(\vec{q}-\vec{k})
\widetilde{V}(\vec{k}')\widetilde{V}(-\vec{q}-\vec{k}')\right> \nonumber \\
& & +{\cal O}(W^3). \nonumber
\end{eqnarray}
Diagrammatically, the second-order
contributions may be represented as in Fig. \ref{Wdiag}. In the Appendix
we compute the above averages in $D=2$. Note that although the
random potential is assumed uncorrelated in space, the screened potential
develops correlations and $\widetilde{W}(\vec{q})$ becomes a non-trivial
function of the wave-vector. For the low-energy states 
one expects the localisation properties to be determined
by $\widetilde{W}(\vec{q})$ at small $\vec{q}$,
so we focus on the limit $\vec{q}\rightarrow 0$ and denote
$\widetilde{W}(\vec{q}\rightarrow 0)= \widetilde{W}$. To the second order in
$W$ in the limit $\Omega\rightarrow 0$ and in
$D=2$ one then finds (see the Appendix for details):
\begin{eqnarray}
\label{all}
\widetilde{W} & = & \frac{W}{\lambda^2c^2}\Omega^2 +
\left(\frac{W}{\lambda^2c^2}\right)^2\Omega^2\Bigg\{\frac{1}{2\pi^5}\left(\left(
\frac{\Lambda}{\Omega}\right)^2 + \frac{32}{\pi}\left(\frac{\Lambda}{\Omega}\right)
\right) \\  \nonumber
& & + \frac{4}{\pi^4}\left(\frac{\Lambda}{\Omega}\right) + {\cal
O}\left(\ln\left(\frac{\Lambda}{\Omega}\right)\right)\Bigg\} + O(W^3), 
\end{eqnarray}
where the constant $c=1/(8\pi)$, and $\Lambda$ is the ultraviolet cutoff
implicit in (\ref{WDE}).

The last equation is our central result, and  
several remarks are in order.
To the first order in $W$, one finds that as $\Omega\rightarrow 0$, 
$\widetilde{W}\rightarrow 0$, which one may be tempted to
interpret as a sign of delocalisation of the ground state. This
is a consequence of the screening of the random potential by the medium,
which to the zeroth order in disorder is pure and thus screens perfectly
at $q=0$. Also, recognising the combination $\widetilde{W}/\Omega^2 $ as
a dimensionless measure of screened
disorder, to the lowest order Eq. (\ref{all}) agrees with the
Harris criterion: disorder is marginal in $D=2$. The fate of disorder is
therefore determined by the higher-order terms in the expansion. 
To the second order in disorder we find that
\begin{equation}
\widetilde{W} 
\rightarrow \frac{\Lambda^2}{2\pi^5}\left(\frac{W}{\lambda^2c^2}\right)^2,
\hspace{1em}{\rm as}\hspace{1em}\Omega \rightarrow 0, 
\end{equation}
i.e. goes to a non-universal finite constant as the gap decreases.
If the bare disorder is weak the screened disorder will be even weaker,
but always finite. The consequence is that the ground state
and the excited states in $D=2$ should remain localised \cite{abrahams},
so that our qualitative argument from the introduction would
imply that the gap can not close. This is in accordance with the
direct numerical solution at strong disorder to which we turn next.

\section{Numerical solution}

We begin by introducing a discrete version of our theory where the
continuous variable $\vec{x}$ is replaced by a lattice-site index $i$ on a quadratic
lattice of linear size $L$. The kinetic energy term $\nabla^2$
is replaced by the nearest-neighbour hopping measured by $t$,
the random potential is chosen from a uniform
distribution of width $W$ and the interaction strength is given by $\lambda$.
In our calculations we set $W/t = 4$ and $\lambda/t = 8$, which
corresponds to strong disorder and interactions.  After the 
integration over frequency, the self-consistent Eq. (\ref{SP1}) becomes
\begin{equation}
\label{SP1l}
\chi_i = \lambda\sum_{\alpha=1}^{N}\frac{\left|\phi_\alpha(i)\right|^2}{\sqrt{\varepsilon_\alpha}},
\end{equation}
in the $\left\{\phi_\alpha(i)\right\}$ basis where these wave functions are now
eigenvectors of the matrix
\begin{equation}
\label{SP3'}
\sum_j\left\{-t_{ij} + \left(V_i + \chi_i - \mu\right)\delta_{i,j}\right\}\phi_\alpha(j)
= \varepsilon_\alpha\phi_\alpha(i)
\end{equation}
where $t_{ij}$ is non-zero for nearest-neighbour $i, j$ only.

We solve the set of $L\times L$ equations using the
 Newton-Raphson algorithm. We gradually increase the chemical potential
$\mu$, using the last found solution as the initial guess at the next $\mu$.
 Finally, we average
over many disorder configurations. Of course, for finite $L$ the gap is always
finite, so to infer the result in the
thermodynamic limit we make the standard finite-size
scaling {\it ansatz} for the average ground state energy
\begin{equation}
\label{e0ansatz}
\varepsilon_0 = L^{-z}F\left[L^{\frac{1}{\nu}}(\mu-\mu_c)\right],
\end{equation}
where $z$ is the {\it dynamical} critical exponent, $\nu$ is the correlation length exponent,
and $\mu_c$ is the critical point in the thermodynamic limit; $F(x)$ is a universal
scaling function.  The values of $z$ and
$\mu_c$ are determined by scaling the $\varepsilon_0$-axis until all curves cross at
a single point.  The exponent $\nu$ is found by scaling the
$\mu$-axis so that
a reasonable collapse of all the data onto a single curve is achieved.

\begin{figure}[h]
\begin{center}
\includegraphics[scale=0.5]{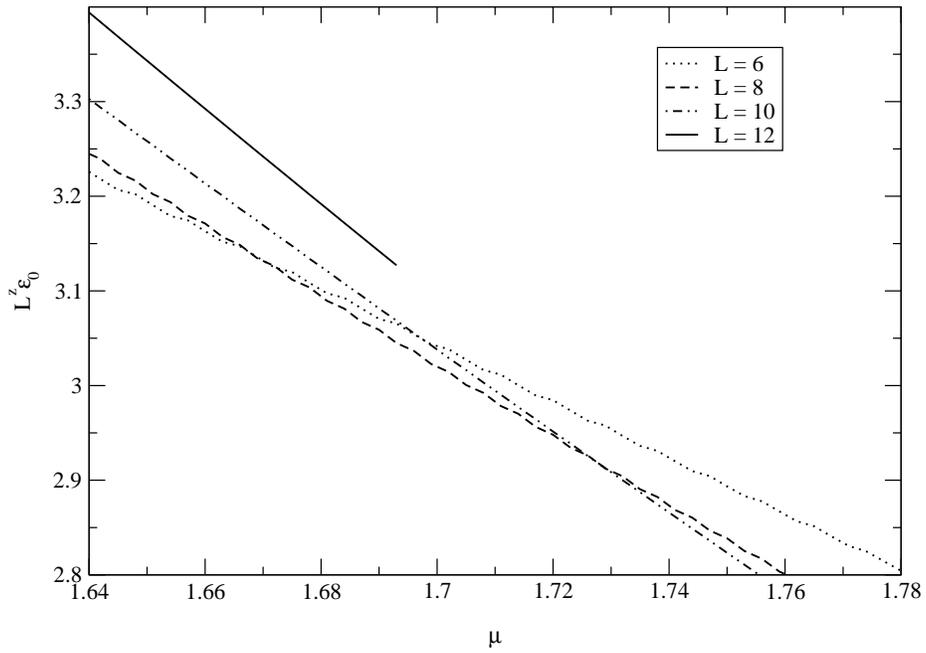}
\end{center}
\caption{Finite size scaling attempt of the ground state energy $\varepsilon_0$ with $z =
0.9$ demonstrating the lack of a transition in our numerical calculations.
The disorder averaging was done over 500 configurations for $L=6$, 1200 for $L=8$,
1000 for $L=10$ and 1000 for $L=12$.}
\label{e0scale}
\end{figure}

Such an attempt of finite size scaling of our data
is shown in Fig. \ref{e0scale}
for systems of linear size $L=6,8,10,12$. We display the result for  
the value $z=0.9$, but the picture remains qualitatively the same
for all $0.5<z<1.0$. The gap continuously 
decreases with $\mu$, but the failure of the finite size scaling
suggests it does not vanish  in the thermodynamic limit.

\begin{figure}[h]
\begin{center}
\includegraphics[scale=0.5]{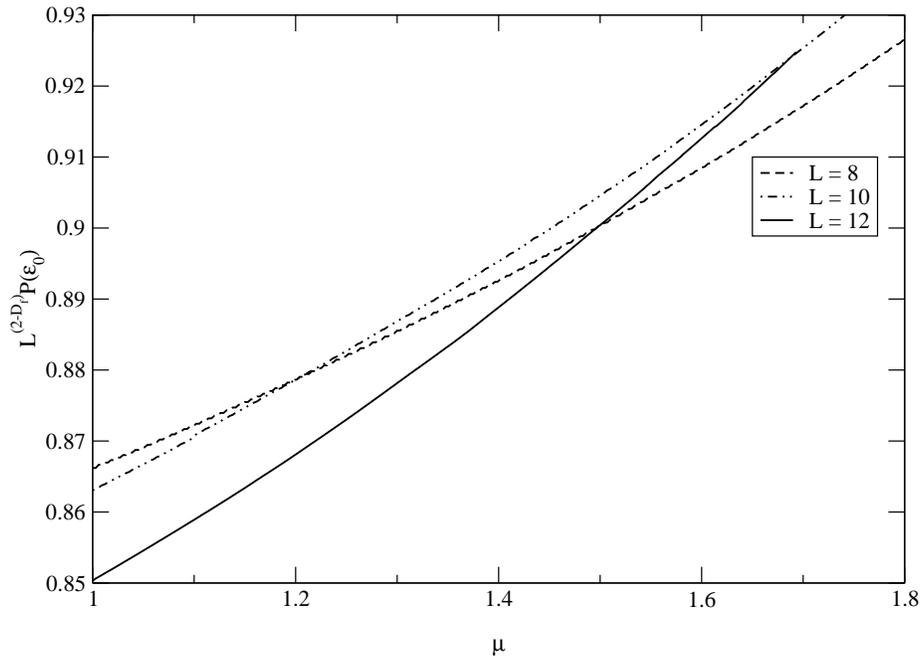}
\end{center}
\caption{Finite size scaling attempt of the ground state participation ratio with $D_f =
0.5$.  Again, the inability to cross these curves at a common point indicates the lack of
the transition.  Disorder averaging was done over the same configurations as in
Fig. \ref{e0scale}.}
\label{PRscale}
\end{figure}

We have also argued that at the point of collapse of the gap,
the ground state would be expected to become delocalised. A useful
measure of the degree of localisation of the  wavefunctions
at a given energy is provided by the participation ratio
\begin{equation}
\label{PR}
P(\varepsilon) = \sum_\alpha\frac{\delta(\varepsilon_\alpha -
\varepsilon)}{L^2\displaystyle\sum_{i=1}^{L^2}\left|\phi_\alpha(i)\right|^4},
\end{equation}
which is proportional to $1/L^2$ for the localised states
and approaches a constant for the extended ones. In the critical region, one
expects the participation ratio to assume a similar finite-size scaling form:
\begin{equation}
\label{PRansatz}
P(\varepsilon_0) = L^{-(D-D_f)}\Phi\left[L^{\frac{1}{\nu}}(\mu-\mu_c)\right],
\end{equation}
where $D_f$ is the fractal dimension of the ground state
wavefunction and $\Phi(x)$ another scaling function.
Our data for the participation ratio are shown in Fig. \ref{PRscale}
for the sizes $L=8, 10,
12$.  Again, attempts to find the common crossing point by tuning 
$D_f$ fail. We see that the participation ratio of the ground state grows
as $\mu$ is increased, but conclude that the ground state nevertheless seems
to remain localised. This is consistent with the data for the ground state
energy.

Our Newton-Raphson algorithm has difficulties converging as
$\mu$ is increased and the problem becomes more non-linear.
It is possible we simply have not been able to reach the critical point
in our numerical calculation.
When taken together with the weak-disorder expansion and the physical
arguments, however, we believe a more likely interpretation is that
there is no SF phase in the model.   

\section{Conclusion}

To summarise, we studied the large-N limit of the commensurate dirty boson
theory, and argued that at weak disorder screening does not delocalise the ground state,
and consequently, that there is no MI--SF transition in $D=2$. Numerical
results for the ground state energy and the participation ratio
that support this conclusion were provided.

  Our conclusion agrees with the results of Kim and Wen \cite{kim}
  who found that disorder is always relevant for $D\geq 2$ and could not find
  any stable critical points within their renormalisation scheme. The latter
  point may in principle be interpreted in three ways: as a failure of the
  renormalisation procedure, as that the transition is discontinuous, or that
  there is no transition. Our findings support the third conclusion. On the
  other hand, we are in disagreement with the recent study of Hastings
  \cite{hastings}, who considered the closely related random spherical model,
  and found the disorder to be marginally {\it irrelevant} in $D=2$.  At the
  moment we do not fully understand what is the resolution of this
  disagreement, nor how the ground state becomes extended
  in Hastings' theory.

While we were mostly concerned with $D=2$, the same perturbative
procedure can be repeated in $D=3$. We found that the same diagram in
Fig. \ref{Wdiag}(e)
that led to the
finite term for $\widetilde{W}$ in $D=3$ vanishes logarithmically as the gap
decreases. More importantly, in $D=3$ the Anderson localisation problem
allows a mobility edge, so the screened disorder need not go all
the way to zero for the ground state to delocalise. We would therefore
expect that the theory (1) would have a MI--SF transition in $D=3$, as
apparently has been found in earlier numerical calculations \cite{hartman}.

An important question is what our considerations imply for the
physical cases $N=1,2,3$ mentioned in the introduction. We believe that
in $D=2$, for $N=2$ the theory (1) does have a transition and which is in the
BG--SF universality class. This has been found in the dual theory for
the commensurate dirty-bosons \cite{herbut}, in both $D=1$ and $D=2$,
and in detailed numerical calculations \cite{wallin, rapsch}.
The BG--SF transition
is best understood in terms of disorder-induced proliferation of
topological defects, and thus is very specific to having a complex
($N=2$) order parameter. The same topological mechanism will not apply
to the case of a random quantum ferromagnet $N=3$, and we conjecture
that for $N=3$ there may not be a gapless phase in $D=2$. On the same grounds,
we expect that for the Ising case $N=1$ the transition again will exist
\cite{dfisher}.

Finally, we note the similarity between our problem and the problem
of interacting disordered fermions in $D=2$ \cite{abrahams1}. In the large-N
limit the metallic phase in the fermionic problem would correspond to an
extended state at the Fermi level, as opposed to the extended ground state
in our problem. Nevertheless, one can show \cite{herbut2} that already to the
lowest order in disorder, screened disorder remains finite,
and thus the state should remain localised. We would therefore expect that
the fermionic version of the action (1) also should have only the localised
phase in $D=2$, at least in the large-N limit.

\section{Acknowledgment}

This research has been supported by NSERC of Canada and the Research
Corporation.

\section{Appendix}

\begin{figure}[h]
\begin{center}
\includegraphics{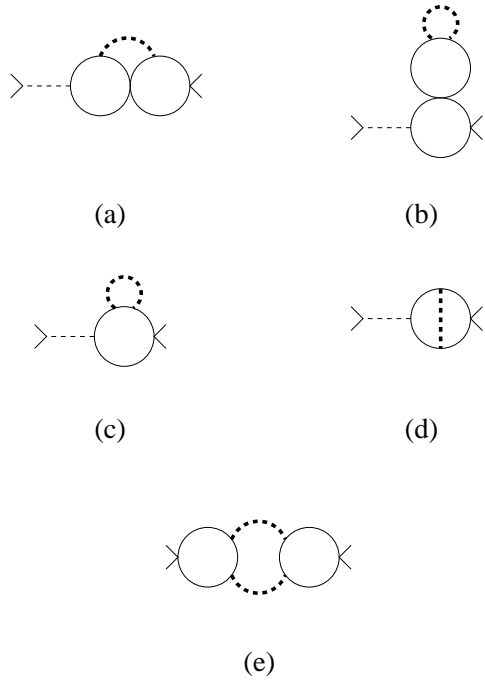}
\end{center}
\caption{Diagrams corresponding to the second order terms in the
expansion (\ref{WDEW}).}
\label{Wdiag}
\end{figure}

In this appendix, we provide the details of the calculations leading up to our main
analytic result (\ref{all}). We begin by calculating the integrals
(\ref{Pi})--(\ref{I2}).  Using the standard Feynman parameters \cite{zinn}
the integrals can be rewritten as 
\begin{eqnarray}
\label{APi}
\Pi(\vec{q}) & = & \frac{\Gamma\left(\frac{3-D}{2}\right)}{(4\pi)^{\frac{D+1}{2}}}
\displaystyle\int_0^1{\rm d}t\frac{1}{\left[t(1-t)q^2 + \Omega^2\right]^\frac{3-D}{2}}
\nonumber \\
& \rightarrow & \left\{
\begin{array}{ll}
\frac{c}{\Omega}\left[\pi\left(\frac{\Omega}{q}\right) - 4\left(\frac{\Omega}{q}\right)^2
+ {\cal O}\left(\left(\frac{\Omega}{q}\right)^3\right)\right], & \frac{q}{\Omega}
\rightarrow \infty \\
\frac{c}{\Omega}, &
q \rightarrow 0,\\
\end{array}
\right.
\end{eqnarray}

\begin{eqnarray}
\label{AI1}
I_1(\vec{k}, 0) & = & \frac{1}{2}\frac{\Gamma(\frac{5-D}{2})}{(4\pi)^\frac{D+1}{2}}
\displaystyle\int_0^1{\rm d}t\frac{1}{\left[t(1-t)k^2 + \Omega^2\right]^\frac{5-D}{2}}
\nonumber \\
& \rightarrow &
\left\{
\begin{array}{ll}
\frac{c}{4\Omega^3}\left[4\left(\frac{\Omega}{k}\right)^2 -
16\left(\frac{\Omega}{k}\right)^4 + {\cal O}\left(\left(\frac{\Omega}{k}\right)^6\right)
\right],
& \frac{k}{\Omega} \rightarrow \infty \\
\frac{c}{4\Omega^3}, & k \rightarrow 0. \\
\end{array}
\right.
\end{eqnarray}
where we assumed $D=2$ in evaluating the limits. 
The diagrams in Figs. \ref{Wdiag}(c),(d) will require the evaluation of
the following two limits of $I_2$:
\begin{eqnarray}
\label{AI21}
I_2(\vec{k}, 0, 0) & = & \frac{\Gamma\left(\frac{7-D}{2}\right)}{(4\pi)^\frac{D+1}{2}}
\displaystyle\int_0^1{\rm d}t\frac{t(1-t)}{\left[t(1-t)k^2 +
\Omega^2\right]^\frac{7-D}{2}} \nonumber \\
& \rightarrow &
\begin{array}{ll}
\frac{3c}{4\Omega^5}\left[\frac{8}{3}\left(\frac{\Omega}{k}\right)^4 +
{\cal O}\left(\left(\frac{\Omega}{k}\right)^6\right)\right], & \frac{k}{\Omega}
\rightarrow \infty \\
\end{array}
\end{eqnarray}
\begin{eqnarray}
\label{AI2}
I_2(0, \vec{l}, 0) & = & \frac{1}{2}\frac{\Gamma\left(\frac{7-D}{2}\right)}
{\left(4\pi\right)^\frac{D+1}{2}}\displaystyle\int_0^1{\rm d}t\frac{(1-t)^2}
{\left[t(1-t)l^2 + \Omega^2\right]^\frac{7-D}{2}} \nonumber \\
& \rightarrow &
\begin{array}{ll}
\frac{3c}{8\Omega^5}\left[\frac{2}{3}\left(\frac{\Omega}{l}\right)^2 + {\cal
O}\left(\left(\frac{\Omega}{l}\right)^6\right)\right], & \frac{l}{\Omega} \rightarrow
\infty \\
\end{array}
\end{eqnarray}
where the limits $\frac{p}{\Omega} \rightarrow \infty$ and $p \rightarrow 0$ are
taken with fixed $p$ and $\Omega$, respectively,
and in $D = 2$.  We also define
$c \equiv \displaystyle\frac{\Gamma(\frac{1}{2})}{(4\pi)^{\frac{3}{2}}} =
\displaystyle\frac{1}{8\pi}$.

We can now evaluate the series (\ref{WDEW}) term-by-term.  From the first order term, in the
limit $\vec{q} \rightarrow 0$, we get
\begin{eqnarray}
\label{AOW}
\widetilde{W}_1(\vec{q}\rightarrow 0) & \equiv & \frac{W}
{\left(1+\lambda\Pi(0)\right)^2}
\nonumber \\
& = & \frac{W}{\lambda^2c^2}\Omega^2
\end{eqnarray}
The contributions of order ${\cal O}(W^2)$ are shown diagrammatically in Fig.
\ref{Wdiag}.
Referring to this figure, we label the corresponding terms generated in the
expansion
accordingly.  To illustrate our procedure, we will explicitly calculate the
diagram shown
in Fig. \ref{Wdiag}(e) arising from the final term in (\ref{WDEW}).
This term is
\begin{equation}
\label{AOW22e1}
\widetilde{W}_{2(e)}(\vec{q}) = \frac{\lambda^2}{\left\{1 + \lambda\Pi(\vec{q})\right\}^2}
\int{\rm d}\vec{k}{\rm d}\vec{k}' I_1(\vec{k}, \vec{q})I_1(\vec{k}',
-\vec{q})\left<\widetilde{V}(\vec{k})\widetilde{V}(\vec{q}-\vec{k})\widetilde{V}
(\vec{k}')\widetilde{V}(-\vec{q}-\vec{k}')\right>,
\end{equation}
where
\begin{equation}
\label{AOW22e2}
\left<\widetilde{V}(\vec{k})\widetilde{V}(\vec{q}-\vec{k})\widetilde{V}
(\vec{k}')\widetilde{V}(-\vec{q}-\vec{k}')\right> =
2\left<\widetilde{V}(\vec{k})\widetilde{V}(\vec{k}')\right>\left<\widetilde{V}
(\vec{q}-\vec{k})\widetilde{V}(-\vec{q}-\vec{k}')\right>
\end{equation}
are the contractions which contribute for $\vec{q} \neq 0$.  Using the definition
\begin{equation}
\label{AOW22e3}
\widetilde{W}(\vec{q})\delta(r) \equiv \left<\widetilde{V}(\vec{q})\widetilde{V}
(-\vec{q}+\vec{r})\right>,
\end{equation}
and integrating over $\vec{k}'$, (\ref{AOW22e1}) becomes
\begin{eqnarray}
\label{AOW22e4}
\widetilde{W}_{2(e)}(\vec{q}) & = &
\frac{2\lambda^2}{\left\{1+\lambda\Pi(\vec{q})\right\}^2}\int{\rm d}\vec{k}I_1(\vec{k},
\vec{q})I_1(-\vec{k}, -\vec{q})\widetilde{W}(\vec{k})\widetilde{W}(\vec{q}-\vec{k})
\nonumber \\
& \rightarrow &
\begin{array}{ll}
\displaystyle\frac{2\lambda^2}{\left\{1+\lambda\Pi(0)\right\}^2}
\int{\rm d}\vec{k}I_1^2(\vec{k}, 0)
\widetilde{W}^2(\vec{k}), & \vec{q} \rightarrow 0.
\end{array}
\end{eqnarray}
We now replace $\widetilde{W}(\vec{k})$ in (\ref{AOW22e4}) to first order in $W$ to get
\begin{eqnarray}
\label{AOW22e5}
\widetilde{W}_{2(e)}(\vec{q} \rightarrow 0) & = &
\frac{2\lambda^2W^2}{\left\{1+\lambda\Pi(0)\right\}^2}\int{\rm d}
\vec{k}\frac{I_1^2(\vec{k}, 0)}{\left\{1+\lambda\Pi(\vec{k})\right\}^4} \nonumber \\
& = & \frac{1}{\pi^5}\left(\frac{W}{\lambda^2c^2}\right)^2\Omega^2
\int_0^{\frac{\Lambda}{\Omega}}x{\rm d}x\frac{\left[1 - 4\left(\frac{1}{x^2}\right) + {\cal
O}\left(\frac{1}{x^4}\right)\right]}{\left[1 - \frac{4}{\pi}\left(\frac{1}{x}\right) +
{\cal O}\left(\frac{1}{x^2}\right)\right]},
\end{eqnarray}
where the last line follows from substituting (\ref{APi}) and
(\ref{AI1}) into the previous line and making the change of variable
$x = \displaystyle\frac{k}{\Omega}$;
$\Lambda$ is the usual ultraviolet cutoff imposed by the lattice.  Expanding the
denominator in (\ref{AOW22e5}) and integrating over $x$ now yields the result
\begin{equation}
\label{AOW22efinal}
\widetilde{W}_{2(e)}(\vec{q}\rightarrow 0) = \frac{1}{2\pi^5}\left(\frac{W}{\lambda^2c^2}
\right)\Omega^2\left[\left(\frac{\Lambda}{\Omega}\right)^2 +
\frac{32}{\pi}\left(\frac{\Lambda}{\Omega}\right) + {\cal
O}\left(\ln\left(\frac{\Lambda}{\Omega}\right)\right)\right].
\end{equation}
It is important to note that (\ref{AOW22efinal})
does not vanish as $\Omega \rightarrow 0$.

The calculation of the remaining terms now follows in a similar way.
The diagrams arising from the second term on the RHS of (\ref{WDEW}) are
those shown in Figs. \ref{Wdiag}(a),(b).  These give
\begin{equation}
\label{AOW22a}
\widetilde{W}_{2(a)}(\vec{q}\rightarrow 0) = \frac{4}{\pi^4}\left(
\frac{W}{\lambda^2c^2}\right)^2\Omega^2\left[\left(\frac{\Lambda}{\Omega}\right) +
{\cal O}\left(\ln\left(\frac{\Omega}{\Lambda}\right)\right)\right],
\end{equation}
and
\begin{equation}
\label{AOW22b}
\widetilde{W}_{2(b)}(\vec{q}\rightarrow 0) =
\frac{1}{4\pi^3}\left(\frac{W}{\lambda^2c^2}\right)^2\Omega^2\left[
\left(\frac{\Lambda}{\Omega}\right)^2 + \frac{16}{\pi}\left(\frac{\Lambda}{\Omega}\right)
+ {\cal O}\left(\ln\left(\frac{\Lambda}{\Omega}\right)\right)\right].
\end{equation}
The third term on the RHS of (\ref{WDEW}) gives rise to the diagrams \ref{Wdiag}(c),(d).
These give
\begin{equation}
\label{AOW22c}
\widetilde{W}_{2(c)}(\vec{q}\rightarrow 0) =
-\frac{1}{4\pi^3}\left(\frac{W}{\lambda^2c^2}\right)^2\Omega^2\left[\left(
\frac{\Lambda}{\Omega}\right)^2 + \frac{16}{\pi}\left(\frac{\Lambda}{\Omega}\right) + {\cal
O}\left(\ln\left(\frac{\Lambda}{\Omega}\right)\right)\right],
\end{equation}
and
\begin{equation}
\label{AOW22d}
\widetilde{W}_{2(d)}(\vec{q}\rightarrow 0) =
-\frac{2}{\pi^3}\left(\frac{W}{\lambda^2c^2}\right)^2\Omega^2\left[{\cal
O}\left(\ln\left(\frac{\Lambda}{\Omega}\right)\right)\right].
\end{equation}
Note that the two highest order terms in (\ref{AOW22c}) cancel exactly with those in
(\ref{AOW22b}).
Summing the contributions (\ref{AOW}) and (\ref{AOW22efinal})-(\ref{AOW22d}),
we then get the result quoted in (\ref{all}). As mentioned, the second order term that
remains constant when $\Omega \rightarrow 0$ comes entirely from the
diagram \ref{Wdiag}(e).

\pagebreak

\pagebreak

\section{Figures}

\begin{thebibliography}{99}
\bibitem{hertz}{J. A. Hertz, L. Fleishman, and P. W. Anderson, Phys. Rev. Lett. {\bf 43},
942 (1979).}
\bibitem{fisher}{M. P. A. Fisher, P. B. Weichman, G. Grinstein, and D. S. Fisher, Phys.
Rev. B {\bf 40}, 546 (1989).}
\bibitem{herbut}{For a review, see I. F. Herbut, Phys. Rev. B {\bf 57}, 13729 (1998) and
references therein.}
\bibitem{belitz}{D. Belitz and T. Kirkpatrick, Rev. Mod. Phys. {\bf 66}, 261 (1994).}
\bibitem{wallin}{M. Wallin, E. S. Sorensen, S. M. Girvin, and A. P. Young, Phys. Rev.
B {\bf 49}, 12115 (1994).}
\bibitem{herbut1}{I. F. Herbut, Phys. Rev. Lett. {\bf 81}, 3916 (1998);
Phys. Rev. B {\bf 58}, 971 (1998); Phys. Rev. B {\bf 61}, 14723 (2000).}
\bibitem{tu}{Y. Tu and P. B. Weichman, Phys. Rev. Lett. {\bf 73}, 6 (1994).}
\bibitem{remark}{More precisely, for commensurate dirty-bosons the action contains an
additional term linear in the time derivative with a random coefficient with zero
average.  We neglect this additional disorder for simplicity.}
\bibitem{dorogovtsev}{S. N. Dorogovtsev, Phys. Rev. Lett. {\bf 76A}, 169 (1980); D.
Boyanovski and J. L. Cardy, Phys. Rev. B {\bf 26}, 154 (1982); I. D. Lawrie and V. V.
Prudnikov, J. Phys. C {\bf 17}, 1655 (1984).}
\bibitem{sachdev}S. Sachdev,
{{\it Quantum Phase Transitions}, (Cambridge University Press,
Cambridge, 1999).}
\bibitem{kim}{Y. B. Kim and X. G. Wen, Phys. Rev. B {\bf 60}, 9755 (1999).}
\bibitem{hastings}{M. Hastings, Phys. Rev. B {\bf 60}, 9755 (1999).}
\bibitem{zinn}J. Zinn-Justin,
{{\it Quantum Field Theory and Critical Phenomena},
(Cambridge University Press, Cambridge, 1996).}
\bibitem{harris}{A. B. Harris, J. Phys. C {\bf 7}, 1671 (1974).}
\bibitem{abrahams}{E. Abrahams, P. W. Anderson, D. C. Licciardello, and T. V.
Ramakrishnan, Phys. Rev. Lett. {\bf 42}, 673 (1979).}
\bibitem{hartman}{J. W. Hartman and P. B. Weichman, Phys. Rev. Lett. {\bf 74}, 4584
(1995).}
\bibitem{rapsch} S. Rapsch, U. Schollwock, W. Zwerger, Europhys. Lett.
{\bf 46}, 559 (1999) and references therein.
\bibitem{dfisher} D. Fisher, Phys. Rev. B, {\bf 51 }, 6411 (1995).
\bibitem{abrahams1} For a review, see E. Abrahams, S. V. Kravchenko,
and M. P. Sarachik, Rev. Mod. Phys. {\bf 73}, 251 (2001).
\bibitem{herbut2} I. F. Herbut, Phys. Rev. B {\bf 63 }, 113102 (2001). 
\end{thebibliography}
\end{document}